\documentclass[%
 reprint,
showpacs,preprintnumbers,
 amsmath,amssymb,
 aps,
]{revtex4-1}

\usepackage{graphicx}% Include figure files
\usepackage{dcolumn}% Align table columns on decimal point
\usepackage{bm}% bold math
\usepackage{color}
\begin{document}

\preprint{LA-UR-18-20811}

\title{Sensitivity of Proposed Search for Axion-induced Magnetic Field using Optically Pumped Magnetometers}% Force line breaks with \\

\author{P.-H.~Chu}
\email[Email address: ]{pchu@lanl.gov}
\author{L.~D.~Duffy}
\email[Email address: ]{ldd@lanl.gov}
\author{Y.~J.~Kim}
\email[Email address: ]{youngjin@lanl.gov}
\author{I.~M.~Savukov}
\affiliation{Los Alamos National Laboratory, Los Alamos, New Mexico 87545, USA}
\date{\today}


\date{\today}% It is always \today, today,

\begin{abstract}
We investigate the sensitivity of a search for the oscillating current induced by axion dark matter in an external magnetic field using optically pumped magnetometers (OPMs).  This experiment is based upon the LC circuit axion detection concept of Sikivie, Sullivan, and Tanner~\cite{Sikivie:2014}. The modification of Maxwell's equations caused by the axion-photon coupling results in a minute magnetic field oscillating at a frequency equal to the axion mass, in the presence of an external magnetic field. The axion-induced magnetic field could be searched for using an LC circuit amplifier with an OPM, the most sensitive cryogen-free magnetic-field sensor, in a room temperature experiment, avoiding the need for a complicated and expensive cryogenic system. We discuss how an existing magnetic resonance imaging (MRI) experiment can be modified to search for axions in a previously unexplored part of the parameter space.  Our existing detection setup, optimized for MRI, is already sensitive to an axion-photon coupling of $10^{-7}$ GeV$^{-1}$ for an axion mass near $3\times10^{-10}$ eV, which is already limited by astrophysical processes and solar axion searches. 
%While this is %ruled out by limits
%limited by astrophysics and solar axion searches, 
We show that realistic modifications, and optimization of the experiment for axion detection, can %set a new limit on 
probe the axion-photon coupling up to four orders of magnitude beyond the current best limit, for axion masses between $10^{-11}$ eV and $10^{-7}$ eV.
%$10^{-10}$ eV.

\end{abstract}

\pacs{32..Dk, 11.30.Er, 77.22.-d, 14.80.Va,75.85.+t}% PACS, the Physics and Astronomy
\maketitle

%\tableofcontents
%%%%%%%%%%%%%%%%%%%%%%%%%%%%%%%%%%%%%%%%
\section{Introduction}
The dark matter of the Universe presents one of the biggest unsolved mysteries in physics. The existence of dark matter is inferred from its gravitational effects. Observations from cosmology and astrophysics support 
%One of the biggest mysteries in physics is the existence of dark matter in the Universe. Dark matter is a hypothetical type of matter which can generate gravitational effects. Several evidences from cosmological and astronomical observations have shown 
the existence of dark matter, including the cosmic microwave background (CMB) power spectrum~\cite{ade:2015}, cluster and galactic rotation curves~\cite{zwicky:1933, rubin:1983}, gravitational lensing~\cite{walsh:1979,clowe:2006} and large-scale structure formation~\cite{springel:2006}. The evidence converges on a Universe in which dark matter is a significant component, contributing more than 80\% of the total matter content~\cite{cyburt:2004}. %However, the particle nature of this dark matter still remains to be discovered.
%except the gravitational effects, no other properties of dark matter have been discovered so far. The dark matter is a dominant component, contributing more than 80~\% of the matter content of the Universe~\cite{cyburt:2004}. Physicists have hypothesized many candidates of dark matter 
Many particle candidates have been proposed, including weakly interacting massive particles (WIMPs)~\cite{steigman:1985}, axions~\cite{peccei:1977,peccei:19771,wilczek:1977,weinberg:1978}, sterile neutrinos~\cite{kusenko:2009}, and others~\cite{patrignani:2017}. While many experimental efforts have been conducted, the nature of dark matter still remains unknown.  The latest direct detection of nuclear recoils from WIMP-nucleus scattering has reached the cross section $< 10^{-46}~\text{cm}^2$ around the WIMP mass 10--100 GeV~\cite{Akerib:2017,Aprile:2017}, which is very close to the neutrino floor due to the nuclear scattering by MeV solar neutrinos~\cite{Monroe:2007}. %While many experimental efforts have been conducted over the last years, the nature of dark matter still remains unknown.

The axion is a natural consequence of the Peccei-Quinn (PQ) solution to the strong CP problem of quantum chromodynamics (QCD) ~\cite{peccei:1977}, and was subsequently realized to be an excellent candidate for the dark matter of the Universe.  The strong CP problem is the question of why the observable $\bar{\theta}$, which violates the discrete symmetry operation of charge conjugation and parity (CP), is limited to be extremely small.
The latest neutron electric dipole moment (EDM) measurement~\cite{baker:2006,Pendlebury:2015} %suggested that 
limits $\bar{\theta}$
%, only describing the charge-parity ($CP$) violating term in quantum chromodynamics, is extremely small, 
to be less than $10^{-10}$.
The PQ solution promotes the vacuum angle to a dynamical variable, which naturally relaxes to a small value after the associated, global $U(1)$, PQ symmetry is broken~\cite{peccei:1977}.  The axion is the Nambu-Goldstone boson from the spontaneous breaking of the PQ symmetry.  It acquires a very small mass due to instanton effects at the cosmological QCD phase transition~\cite{wilczek:1977,weinberg:1978,Gross:1980}.
%Peccei and Quinn proposed an elegant solution to this problem: a new global symmetry can be added to the Standard Model of particle physics, effectively promote the $\bar{\theta}$ field which becomes  spontaneously broken resulting in a new Goldstone boson~\cite{Goldstone:1962}, the axion~\cite{peccei:19771,weinberg:1978}. 
Due to its massive and non-relativistic properties, the axion is 
%treated as 
a promising candidate for dark matter.

%Axions have been suggested to couple with photons, electrons, and nucleons~\cite{patrignani:2017}. The Axion-photon coupling has played an important role for many existing searches~\cite{Graham:2015}. 
%Recently, the low mass axion has aroused new attention in physics because of cosmology and string theory~\cite{String1, String2}. Some theoretical models have suggested if the inflation happens after the axion process, the axion field becomes homogeneous and lies close to the minimum of its effective potential, which leads for its mass to be much smaller than $10^{-5}$ eV~\cite{Pi:1984}. 
The traditional axion mass window is considered to be in the range from $10^{-6}$ to $10^{-2}$ eV, based on constraints from astrophysics and cosmology.  The lower bound is based on not overproducing dark matter; however it assumes the axion field is initially far from the minimum of its effective potential (see e.g. Refs.~\cite{Kim:1986,Sikivie:2006,Duffy:2009} for detailed discussion).  This is not necessarily true, and an initial position close to the minimum, prior to cosmological inflation, can lead to an axion mass much smaller than $10^{-6}$ eV, and an abundance that meets cosmological bounds~\cite{String1, String2,Pi:1984}. Additionally, 
string theory favors the energy scale at the Planck scale, which can result in very small axion masses~\cite{Svrcek:2006}.

Sikivie, Sullivan and Tanner have proposed searching for these very light axions with an LC circuit coupled to a sensitive magnetometer~\cite{Sikivie:2014}.  We investigate the sensitivity that can be achieved
%we propose a new experimental approach to directly look for a light mass axion below $10^{-7}$~eV 
using a detection system composed of an LC circuit and an optically pumped magnetometer (OPM), operated at ambient temperatures.  The OPM, based on lasers and alkali-metal vapor cells, is the currently most sensitive cryogen-free magnetic sensor reaching femtotesla sensitivity~\cite{OPM_nature}. We will show that we can search for dark matter axions in the mass range from $10^{-11}$ to $10^{-7}$ eV with a sensitivity to the axion-photon coupling up to 4 orders of magnitude beyond the current best limit achieved by the CERN Axion Slolar Telescope (CAST) experiment~\cite{CAST:NatPhys}.  This represents an important search that can be conducted with existing technology, without the complication of a cryogenic system.
%The use of an LC circuit for the direct search for a light mass axion has been recently suggested by the Sikivie's group~\cite{Sikivie:2014}. 

Axions possess model-dependent couplings to photons, electrons, and nucleons ~\cite{patrignani:2017}.  Many existing searches rely on the axion-photon coupling~\cite{Graham:2015}, as does the proposed LC circuit approach~\cite{Sikivie:2014}. We briefly summarize the theoretical background next.

The effective Lagrangian of the axion-photon interaction is 
%written as
\begin{align}
\mathcal{L} = −gaF_{\mu\nu}\tilde{F}^{\mu\nu} \; ,
\end{align}
where $F_{\mu\nu} = \partial_{\mu}A_{\nu}-\partial_{\nu}A_{\mu}$ is the electromagnetic field-strength tensor, $\tilde{F}^{\mu\nu}=−\frac{1}{2}\epsilon^{\mu\nu\alpha\beta}F_{\alpha\beta}$ is its dual, $A_{\mu}$ is the photon field, $a$ is the axion field, and $g$ is the axion-photon coupling. Here we use natural units with $c=\hbar=\mu_0=1$.
The axion-photon coupling leads to the following modified Maxwell equations~\cite{Sikivie:2014}:
\begin{align}
    \vec{\nabla}\cdot\vec{E} =& g\vec{B}\cdot\vec{\nabla}a + \rho_{el},\notag\\
    \vec{\nabla}\times\vec{B}-\frac{\partial\vec{E}}{\partial{t}}=& g\Big(\vec{E}\times\vec{\nabla}a-\vec{B}\frac{\partial a}{\partial t}\Big)+\vec{j}_{el},
    \label{eq:modifiedmaxwell}
\end{align}
where $\rho_{el}$ and $\vec{j}_{el}$ are electric charge and current densities associated with ordinary matter. In a static  magnetic field, $B_0$, axions can induce an electric current density, $\vec{j}_a = -g\vec{B}_0 \dot{a}$, according to Eq.~\ref{eq:modifiedmaxwell}. Here, $\vec{\nabla}a$ is neglected due to the assumption that the axion field is homogeneous. The axion field, $a=a_0\cos{(\omega t)}$, oscillates at a frequency $\omega=m_a$ where $m_a$ is the axion mass. Then $j_a$ can produce a minute oscillating magnetic field $B_a$, perpendicular to the static magnetic field, through $\vec{\nabla}\times \vec{B}_a = \vec{j}_a$. 
%In our approach, this 
When the resonant frequency of the LC circuit is near the axion mass, the current induced in the pickup loop by $B_a$ will be amplified by the circuit and then sensitively measured by an OPM. If the dark matter consists entirely of axions, the dark matter density is equal to 
\begin{align}
    \rho_{DM} = \frac{1}{2}m_a^2 a_0^2
\end{align}
~\cite{Dine:1983} such that the amplitude of the axion field is $a_0 = \sqrt{2\rho_\text{DM}}/{m_a}$ where $\rho_{\text{DM}}\approx0.3~ \text{GeV/cm}^3$~\cite{patrignani:2017}, assuming the dark matter halo is in thermal equilibrium. 

%The same idea can be applied for dark photons~\cite{Arias:2015}. 
After the proposal of Sikivie, Sullivan and Tanner, it was also realized that the same setup without the applied magnetic field could be adapted to search for hidden sector photon dark matter~\cite{Arias:2015,Chaudhuri:2014dla}. 

%%%%%%%%%%%%%%%%%%%%%%%%%%%%%%%%%%%%%%%%%%%%%%%%%%%%%%%%%%
\section{Experimental Method}
\begin{figure}[t]
\centering
\includegraphics[width=3.3in]{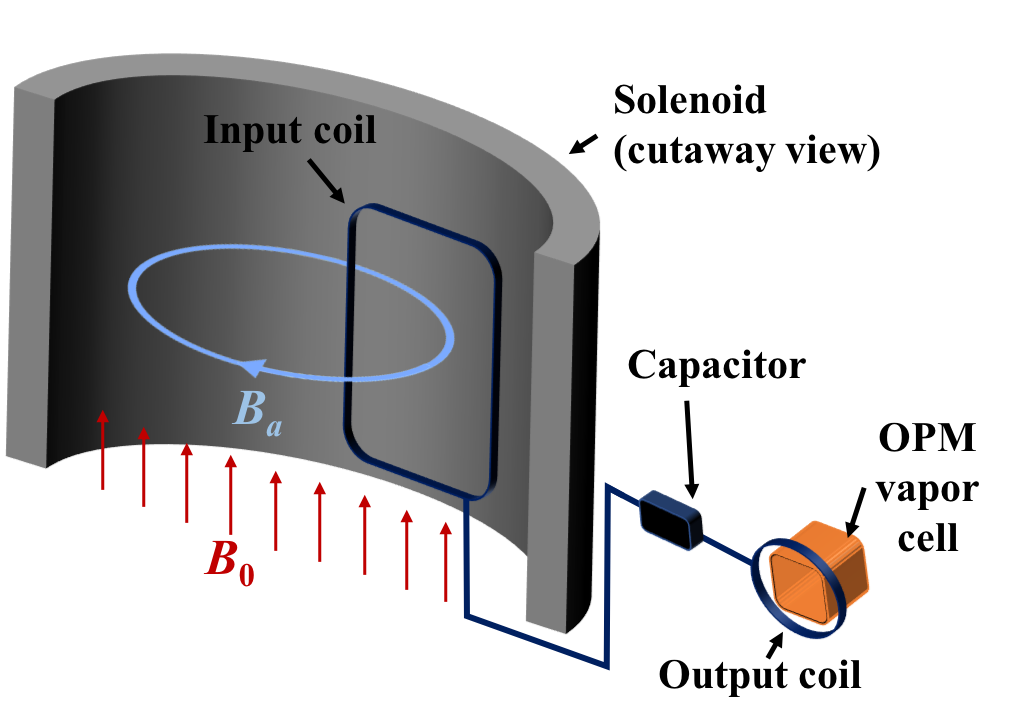}
\caption{\label{Design} Schematic drawing of the axion dark matter search using a LC circuit-OPM detection system. The OPM is simplified as a vapor cell. The output coil is positioned right above the vapor cell in order not to block the laser beams.}
\end{figure}

A schematic drawing of an experimental setup using an OPM is shown in Fig.~\ref{Design}. The detection system is composed of two coils with a capacitor (LC circuit) to amplify the axion-induced AC magnetic signal and an OPM to detect the amplified signal. A solenoid produces a static magnetic field $\vec{B_0}$; then the axion-induced magnetic field $\vec{B_a}$ induces a voltage in a rectangular one-turn input coil (one turn is chosen to minimize the total inductance of the LC circuit), located inside the solenoid, and drives a current through a circular output coil which is detected by an OPM. In Fig.~\ref{Design}, the applied field $\vec{B_0}$ is in the vertical ($z$) direction, resulting in an axion-induced field $\vec{B_a}$ in the azimuthal ($\phi$) direction.  The OPM Zeeman resonance frequency (Larmor frequency) should be matched approximately to the resonance frequency of LC circuit to get the best sensitivity. It should be noted that because $\vec{B_a}$ is azimuthally symmetric for uniform $B_0$, the input coil should cover only one side of the central axis of the solenoid. 

The magnitude of the induced voltage in the input coil is 
%written by 
$V =\omega\Phi_a$, where $\Phi_a$ is the magnetic flux through the input coil and $\omega$ is the operating angular frequency. This voltage drives a current in the LC circuit, 
\begin{align}
I =\frac{\omega\Phi_a}{i\omega(L_{\text{in}}+L_{\text{out}})-i(\omega C)^{-1}+R} \; ,
\label{eq:I_LC}
\end{align}
where  $L_{\text{in}}$ and $L_{\text{out}}$ are the inductances of the input and output coils, respectively, $C$ is the capacitance of the capacitor, and $R$ is the total AC resistance of the LC circuit (lossy capacitor can increase the effective AC resistance of the circuit). Here the inductance of the capacitor is negligible. At the resonance of the LC circuit where $\omega=1/\sqrt{(L_{\text{in}}+L_{\text{out}})C}$ and with 
%the relation of the 
quality factor of the circuit, $Q=\omega(L_{\text{in}}+L_{\text{out}})/R$, Eq.~\ref{eq:I_LC} reduces to 
\begin{align}
I =\frac{Q\Phi_a}{L_{\text{in}}+L_{\text{out}}}.
\end{align}
This current generates a magnetic field in the center of the output coil
\begin{align}
B_{d} =\frac{N_{\text{out}}I}{2r_{\text{out}}}=\frac{N_{\text{out}}Q\Phi_a}{2r_{\text{out}}(L_{\text{in}}+L_{\text{out}})}
\end{align}
where $r_{\text{out}}$ and $N_{\text{out}}$ is the radius and the number of turns of the output coil, respectively. The OPM will measure the field $B_{d}$ with high sensitivity.

%%%%%%%%%%%%%%%%%%%%%%
%%%%%%%%%%%%%%%%%%%%%%%%%%

Using cylindrical coordinates, $(z,\rho,\phi)$, $\vec{B_{0}}=B_0\hat{z}$, %then we have
it follows that
\begin{align}
\vec{B_{a}} =-\frac{g\dot{a}B_0\rho}{2}\hat{\phi},
\label{eq:ba_in}
\end{align}
and the magnetic flux through the input coil is
\begin{align}
\Phi_{a}=\int \vec{B_{a}}\cdot {d\vec{A}}=-\text{V}_{\text{in}}g\dot{a}B_0
\label{eq:phi_in}
\end{align}
where $\text{V}_{\text{in}}=l_{\text{in}}r_{\text{in}}^2 /4$, $l_{\text{in}}$ and $r_{\text{in}}$ is the vertical and horizontal length of the input coil, respectively. Substituting Eq.~\ref{eq:phi_in} to Eq.~\ref{eq:ba_in}, we have the magnitude of the field $B_d$
\begin{align}
    B_d = \frac{N_{\text{out}} Q}{2r_{\text{out}}(L_{\text{in}}+L_{\text{out}})}\text{V}_{\text{in}} g \sqrt{2\rho_{DM}} B_0 
    \label{eq:bd_out}
\end{align}
where we used the relation between the time derivative of the axion field and the axion density, $\dot{a}^2=2\rho_{DM}$. Note that Eq.~\ref{eq:bd_out} is equivalent to Eq.~14 of Ref.~\cite{Sikivie:2014}.

In addition to the field noise of the OPM, $\delta B_{\text{OPM}}$, the dominant source of magnetic noise in the detection system is the Johnson noise in the LC circuit, $\delta V_{\text{J}}$, which is the combined noise from two coils:
\begin{align}
\delta V_{\text{J}}&=\sqrt{4k_B T_{\text{in}}R_{\text{in}}+4k_B T_{\text{out}}R_{\text{out}}}\notag\\
&=\sqrt{4k_B T_{\text{out}}R_{\text{out}}\Big(1+\frac{T_{\text{in}}R_{\text{in}}}{T_{\text{out}}R_{\text{out}}}\Big)}
\end{align}
where $k_{B}$ is the Boltzmann constant, $T_{\text{in(out)}}$ is the temperature of input (output) coil, and $R_{\text{in (out)}}$ is the resistance of the input (output) coil. 
Here $R_{\text{out}}=\rho_{\text{out}}l_{\text{out}}/A_{\text{wire}}$ where $\rho_{\text{out}}$ and $A_{\text{wire}}$ is the resistivity and the cross-section area of the wire of the output coil, respectively, and $l_{\text{out}}=2\pi r_{\text{out}}N_{\text{out}}$ is the total length of the output coil. Note that this simplified calculation valid for low frequency has to be modified to take into account the losses from skin-depth and proximity effects. Litz wire can be used to reduce AC resistance closer to the DC resistance value. We propose to operate the detection system at room temperature, $T_{\text{in}}=T_{\text{out}}$, giving 
\begin{align}
\delta V_{\text{J}}=\sqrt{\frac{8\pi k_B T_{\text{out}}\rho_{\text{out}}r_{\text{out}}N_{\text{out}}}{A_{\text{wire}}}}\sqrt{1+R_{\text{in}}/R_{\text{out}}}.
\end{align}
The Johnson noise can be converted to the magnetic noise, $\delta B_{J}$, by Faraday's law:
\begin{align}
    %\delta V_J &= \frac{\partial \Phi}{\partial t}= \omega\Phi = \omega (\delta B_{J})N_{out} A_{out}\\
    \delta B_{J}&=\frac{\delta V_{J}}{\omega N_{\text{out}}A_{\text{out}}}\notag\\
    &=\frac{1}{\omega r_{\text{out}}}\sqrt{\frac{8k_B T_{\text{out}}\rho_{\text{out}}}{\pi r_{\text{out}}N_{\text{out}}A_{\text{wire}}}}\sqrt{1+R_{\text{in}}/R_{\text{out}}}
    \label{eq:deltaBJ}
\end{align}
where $A_{\text{out}}$ is the cross-section area of the output coil. This equation shows that the magnetic Johnson noise decreases at higher frequency. The total magnetic noise of the detection system is given by $\delta B_{d}=\sqrt{\delta B_{J}^2 +\delta B_{\text{OPM}}^2}$. 

Other possible sources of noise are either very small or can be highly suppressed. The ambient high-frequency noise can be eliminated by using a Faraday cage for its electrical component and a radio-frequency (RF) shield for its electro-magnetic component. The noise from the high-field solenoid might be low because the direction of its field is perpendicular to the sensitive direction of the input coil and it drops quickly with frequency. The noise due to non-orthogonality between the input coil and the field $B_0$ will be suppressed by a small deviation from 90 degrees. Acoustic or thermal expansion geometrical effects are also very small above kHz frequencies. Also, a gradiometer-type input coil can be used to eliminate the common noise: in Fig.~\ref{Design}, another one-turn rectangular coil locates next to the input coil to cover the other side of the central axis of the solenoid where $B_a$ is in opposite direction (a first-order planar gradiometer), which doubles the $B_a$ signal, but cancels out the common magnetic noise. This configuration will double the inductance of the input coil, resulting in different optimal experimental parameters from those in the configuration in Fig.~\ref{Design}. 

\subsection{Preliminary Study}
\begin{figure}[t]
\centering
\includegraphics[width=3.3in]{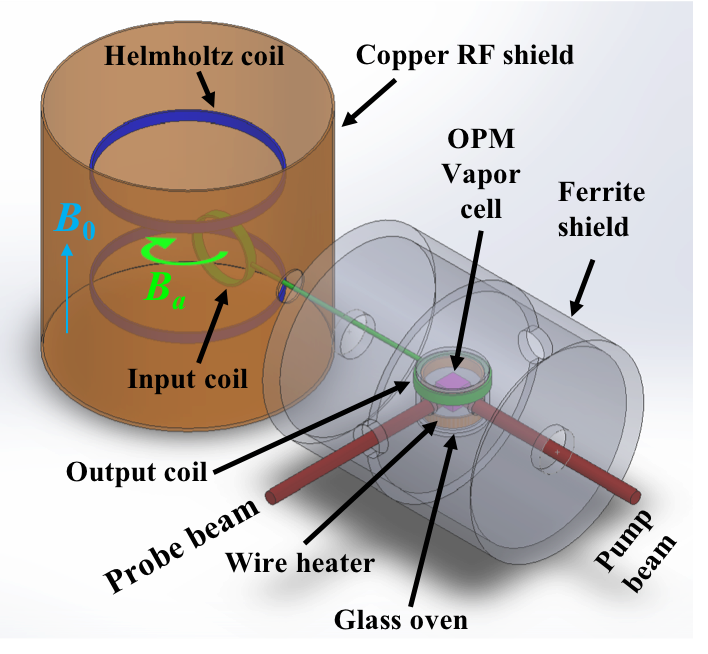}
\caption{\label{System} Schematic diagram of the preliminary detection system composed of a RF OPM and a LC circuit. The input coil is located inside a copper RF shield to reduce the ambient high frequency noise and the output coil and the OPM are located inside a ferrite shield to reach high field sensitivity of the OPM.}
\end{figure}
As a preliminary study for the proposed axion search experiment, we investigated the sensitivity of a detection system composed of a RF OPM constructed at Los Alamos National Laboratory (LANL)~\cite{RF_LANL} and a LC circuit as shown in Fig.~\ref{System}. This system was originally developed for its application to magnetic resonance imaging (MRI) detection~\cite{RF_LANL}; hence its experimental parameters were optimized for MRI experiments. A time-varying magnetic field $B_a$, induced by the axion when an external static magnetic field $B_0$ is applied by the Helmholtz coil, produces a voltage in the input coil (150 turns, a 7.5~cm diameter, a 0.83~mm copper wire diameter, a 493~$\mu$H inductance) located inside a copper RF shield and drives a current through the output coil (40 turns, a 5.5~cm diameter, a 0.25~mm copper wire diameter, a 344~$\mu$H inductance) which is detected by the RF OPM. The OPM and the output coil are located inside a low-noise (much lower than fT/$\sqrt{\text{Hz}}$~\cite{SR_OPM}) cylindrical ferrite shield (of 15~cm length and 10~cm diameter) to reduce the effects of the Earth's field, the external static fields, field gradients, and magnetic noise on the OPM. A capacitor added to the circuit increases the efficiency of flux transfer and selects a desired resonance frequency of the circuit, here 80~kHz. 

The RF OPM consists of a 1~cm cubic potassium (K) vapor cell, a pump and a probe laser beams, optics of mirrors, lenses, a polarizer, a beam splitter, and a quarter wave plate (not shown), and two photodiodes (not shown). The circularly polarized pump beam is used to polarize K atomic spins and the linearly polarized probe beam is used to read out the state of the spins. The pump and probe beams intersect inside the vapor cell at 90$^{\circ}$, which establishes the active volume of the OPM. The action of the pump beam creates a source of a large number of 100\% polarized electron spins in the vapor cell. The interaction of a weak external magnetic field with the polarized spins leads to changes in the orientation of the spins, which is detected through its effect on the light polarization of the probe beam via the Faraday effect. A RF tunable OPM of this type demonstrated a very high sensitivity, 0.2~fT/$\sqrt{\text{Hz}}$~\cite{RF_OPM}. The OPM Zeeman resonance (Larmor frequency, 7~GHz/T in the case of K atomic spins) was tuned to 80~kHz by a bias magnetic field produced by a coil inside the ferrite shield (not shown). The Helmholtz coil, perpendicular to the input coil, produced $B_0=2$~mT; however in the future, the external magnetic field will be increased to tesla level in order to enhance  the magnetic observable (see Eq.~\ref{eq:bd_out}). 

\begin{figure}[t]
\centering
\includegraphics[width=3.45in]{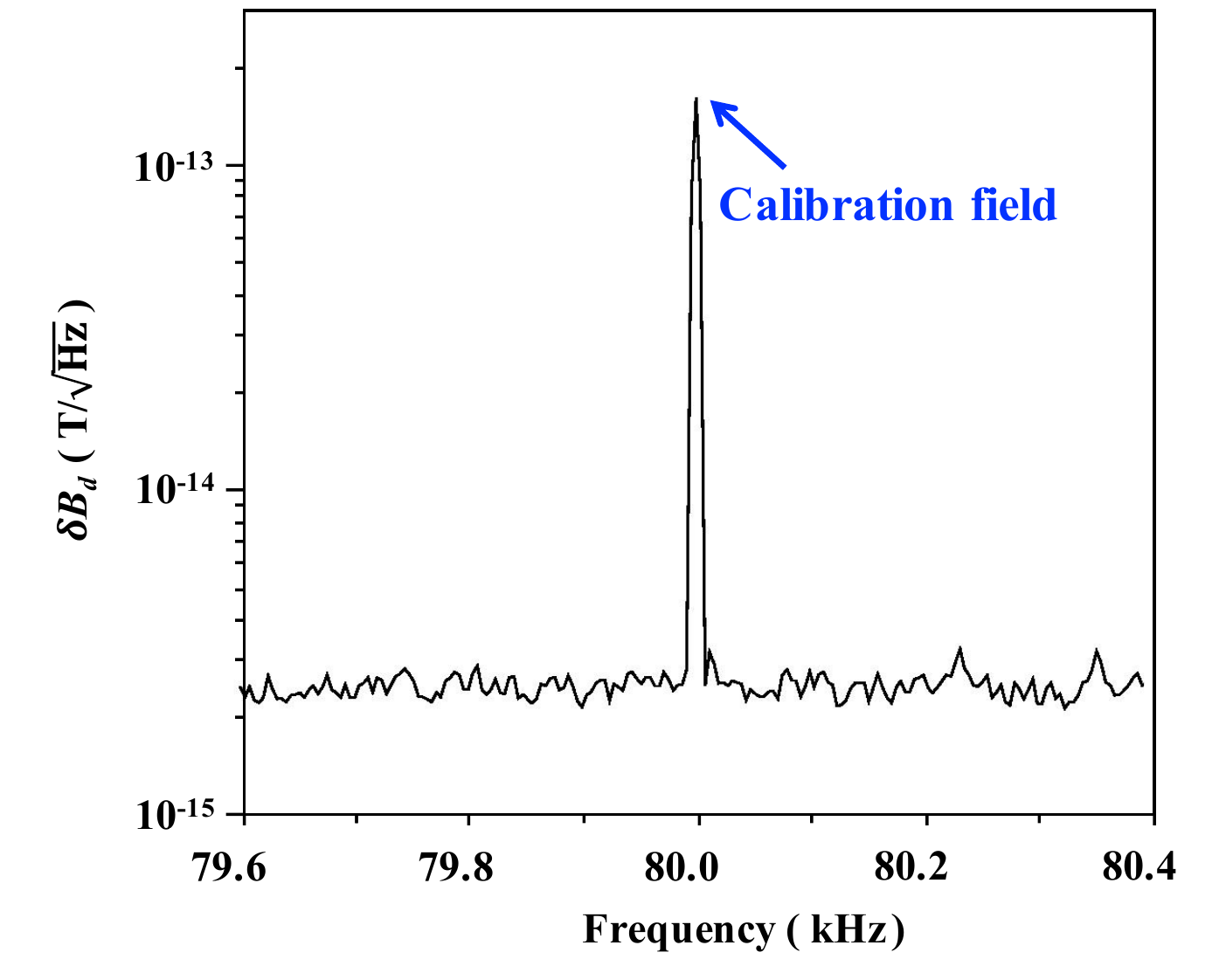}
\caption{\label{Noise} Calibrated magnetic field noise spectrum of the preliminary detection system. The noise near 80~kHz was found to be at the level of 2~fT/$\sqrt{\text{Hz}}$. The large peak at 80~kHz is the calibration field. External RF noise is absent.}
\end{figure}
The total magnetic field noise of the detection system was measured to be $\delta B_{d}=$ 2~fT/$\sqrt{\text{Hz}}$ as shown in Fig.~\ref{Noise}, limited by the magnetic Johnson noise, while the noise of the OPM was measured to be $\delta B_{\text{OPM}}=1$~fT/$\sqrt{\text{Hz}}$. A calibration coil generating a uniform calibration field at 80~kHz was mounted near the vapor cell in order to convert the measured OPM output voltage spectrum into the magnetic field spectrum. The flat noise spectrum indicates that external RF noise was sufficiently suppressed. During the measurements, we found that careful grounding of the detection system suppressed significantly external noise.

\section{Sensitivity Estimate}
Our proposed experiment is based on redesign and optimization of the existing experiment for axion detection. A solenoid-type superconducting magnet with a 1~m bore diameter and a 3~m length, which can generate a 2~T magnetic field, is available in LANL. Based on this magnet and Ref.~\cite{Sikivie:2014}, we propose optimal experimental dimensions of the experimental setup shown in Fig.~\ref{Design}: $B_0=2$~T, $l_{\text{in}}=1$~m, $r_{\text{in}}=0.3$~m, $r_{\text{out}}=1$~cm, and $b=2$~mm copper wire radius for the input and output coils. To maximize the axion-induced magnetic flux through the input coil, $l_{in}$ and $r_{in}$ should match the dimension of the bore of the magnet. The inductance of the input coil is estimated by~\cite{Sikivie:2014}
\begin{align}
L_{\text{in}}\approx& \frac{1}{\pi} l_{\text{in}} ln(r_{\text{in}}/b)=2.0~\mu \text{H}.
\label{eq:L_in}
\end{align}
In addition, the inductance of the output coil is estimated by~\cite{Sikivie:2014}
\begin{align}
L_{\text{out}} &\approx r_{\text{out}} N_{\text{out}}^2 \Big[ln\Big(\frac{8r_{\text{out}}}{b}\Big)-2\Big]\notag\\
&= (2.1\times 10^{-2}\mu \text{H})\times N_{\text{out}}^2.
\label{eq:L_out}
\end{align}
In order to find the optimal value of $N_{\text{out}}$, we substitute Eq.~\ref{eq:L_in} and ~\ref{eq:L_out} to Eq.~\ref{eq:bd_out}
\begin{align}
B_{d} =\frac{Q\Phi_aN_{\text{out}}}{2r_{\text{out}}[2.0~\mu \text{H}+(2.1\times 10^{-2}\mu \text{H})\times N_{\text{out}}^2]}.
\end{align}
The optimal $N_{\text{out}}$ is determined by maximizing $B_{d}$, which happens at $2.0~\mu \text{H}=(2.1\times 10^{-2}\mu \text{H})\times N_{\text{out}}^2$. This results in the optimal $N_{\text{out}}=10$, which leads to $L_{\text{out}}=2.1~\mu$H, thus $L=L_{\text{in}}+L_{\text{out}}=4.1~\mu$H. We assume $Q=400$ which can be reached at kHz and MHz frequencies at room temperature.

\begin{figure}[t]
\centering
\includegraphics[width=3.45in]{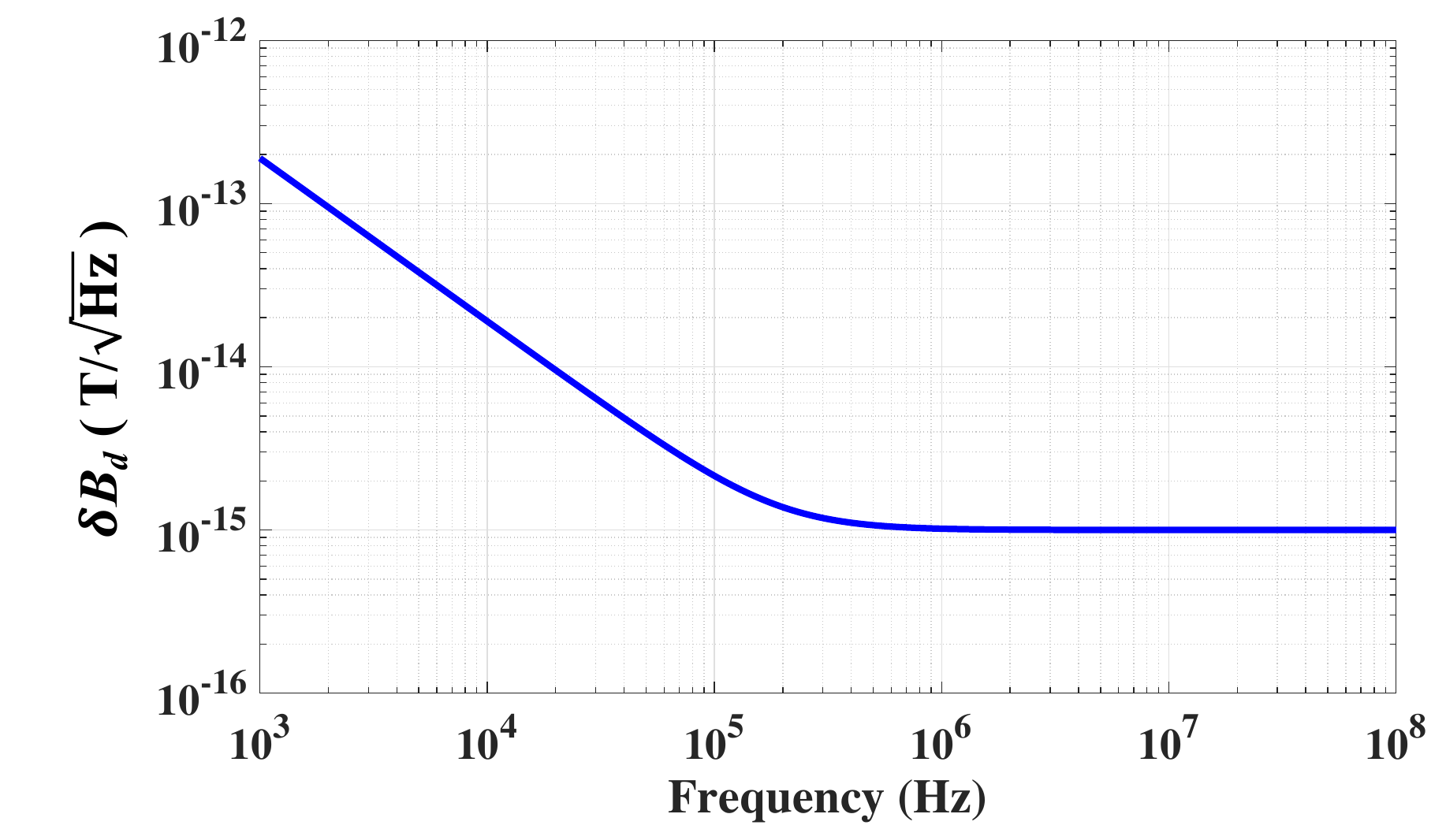}
\caption{\label{Est_Noise} Estimated magnetic field noise of the optimized detection system. The RF OPM with  1~fT/$\sqrt{\text{Hz}}$ at kHz and MHz frequencies will be employed. The noise of the system is limited by the magnetic Johnson noise. }
\end{figure}

We estimate the noise of the optimized detection system  $\delta B_{d}=\sqrt{\delta B_{J}^2 +\delta B_{\text{OPM}}^2}$ based on the experimental parameters. From Eq.~\ref{eq:deltaBJ}, the magnetic Johnson noise is
\begin{align}
    \delta B_{J}=4.3\times10^{-16}~\text{T}/\sqrt{\text{Hz}}\Big(\frac{\text{MHz}}{\nu}\Big)
\end{align}
where $\nu$ is the operating frequency. $R_{\text{in}}$ and $R_{\text{out}}$ is calculated to be 3.5~m$\Omega$ and 0.8~m$\Omega$, respectively. We will use our RF OPM~\cite{RF_LANL} with $\delta B_{\text{OPM}}=1$~fT/$\sqrt{\text{Hz}}$ at high frequencies which we used in the preliminary study. Hence the estimated noise of the system is given by
\begin{align}
    \delta B_{d}=\sqrt{\Big[4.3\times10^{-16}\Big(\frac{\text{MHz}}{\nu}\Big)\Big]^2+(1\times10^{-15})^2}~\text{T}/\sqrt{\text{Hz}}.
\end{align}
Figure~\ref{Est_Noise} shows the estimated noise of the optimized detection system as a function of the frequency. This detection system loses sensitivity at the frequency range below about 50~kHz due to the magnetic Johnson noise, limiting the lower frequency range of this experiment at room temperature. 

In principle, the experimental sensitivity can be enhanced by a long measurement. The bandwidth of the optimized detection system is characterized by quality factor of its LC circuit, $\Delta \nu=\nu/Q$, which leads to the bandwidth of the system $\Delta \nu=2.5\times10^{-3}\nu$.  While our lower $Q$ of 400 means less signal amplification than the high $Q=10000$ of Ref.~\cite{Sikivie:2014}, 
the larger bandwidth means that our system is able to scan the frequency range, and thus the axion mass range, 25 times faster. If it scans a factor of 2 in frequency per year and the duty factor is 30~\% (using the same assumptions as Ref.~\cite{Sikivie:2014} for comparison), the data integration time available at each tune of the LC circuit is $2.5\times10^{4}$~s. Thus, we can partially overcome the sensitivity loss of lower $Q$ by using longer integration times, due to our greater bandwidth.
%Because of the lower $Q$, our detection system can have the longer integration time than Ref.~\cite{Sikivie:2014}, which is one of advantages of our proposed experiment. 
Assuming the field noise of the detection system in $B_0=2~$T is similar to what we estimated (Fig.~\ref{Est_Noise}), the noise of the optimized detection system with the integration time $t=2.5\times10^4$ is given by
\begin{align}
&\delta B_{d}\times(t_ct)^{-1/4}\notag\\
&=1.3\times10^{-16}~\text{T}\Big(\frac{\nu}{\text{MHz}}\Big)^{\frac{1}{4}}\sqrt{0.18\Big(\frac{\text{MHz}}{\nu}\Big)^2+1}
\label{eq:bd_opt}
\end{align}
where $t_c=0.16~$s(MHz/$\nu$) is the signal coherence for the isothermal halo model (see Ref.~\cite{Sikivie:2014} and Eq.~A8 in ~\cite{Budker:2014} for the details). At $\nu=80~$kHz, $\delta B_d=3.6\times 10^{-16}$~T. 

Based on the optimized experimental parameters and Eq.~\ref{eq:bd_opt}, we can estimate the sensitivity of our proposed experiment to the axion-photon coupling $g$ with Eq.~\ref{eq:bd_out},
\begin{align}
g &= \text{SNR}\frac{2r_{\text{out}} L\delta B_d }{Q N_{\text{out}}V_{\text{in}}\sqrt{2\rho_{DM}}B_0}\notag\\
&= \text{SNR}\times\Big(\frac{\delta B_d}{10^{-17}~\text{T}}\Big)\Big(\frac{\text{GeV/cm}^3}{\rho_{DM}}\Big)^{\frac{1}{2}}\Big(\frac{10^3}{Q}\Big)\Big(\frac{L}{\mu \text{H}}\Big)\Big(\frac{\text{T}}{B_0}\Big)\notag\\&\times \Big(\frac{1}{N_{\text{out}}}\Big)\Big(\frac{r_{\text{out}}}{\text{cm}}\Big)\Big(\frac{\text{m}^3}{V_{\text{in}}}\Big)(8\times10^{-18}~\text{GeV}^{-1})
\label{eq:g_sen}
\end{align}
where SNR is the signal-to-noise ratio, taken as 5 for axion detection~\cite{Sikivie:2014}. 

\begin{table}[t]
\centering
\begin{tabular}{c|c|c|c}
\hline
Parameters & Preliminary data & Ref.~\cite{Sikivie:2014} proposal & Our proposal \\
\hline
$\delta B_d$ (T)   & 2.0e-15& 1.5e-17 &3.6e-16  \\
$N_{\text{out}}$    & 40 & 1   &  10              \\
$Q$ &  178 &   10000 &   400            \\
$L$ ($\mu H$) & 837 & 2.6 & 4.1  \\
$r_{\text{out}}$ (cm) & 2.8 & 1 & 1\\
$V_{\text{in}}$ (m$^3$) & 8.25e-3 &0.0225 & 0.0225\\
$B_0$ (T) & 2e-3 & 8 & 2 \\
\hline
$g$ (GeV$^{-1}$) & 8.0e-7 & 1.6e-16 & 6.1e-14 \\
\hline
\end{tabular}
\caption{$\delta B_d$ in preliminary data was experimentally obtained with 1~s integration time while $\delta B_d$ in Sikivie's and our proposed experiment are estimated with $10^3$~s and $2.5\times10^4$~s integration time, respectively. These integration times were determined by the bandwidth of the detection systems. The operating frequency is 80~kHz.}
\label{tab:parameters}
\end{table}
We use Eq.~\ref{eq:g_sen} to compare the sensitivity of our proposed experiment with existing limits. The configurations investigated in Ref.~\cite{Sikivie:2014} are the best possible that can be achieved with existing technology and magnets.  This includes a high-$Q$, superconducting input loop, that is cryogenically cooled to 0.5 mK, and a high-field magnet.  In Table~\ref{tab:parameters}, we compare our sensitivity at 80 kHz to the configuration of Ref.~\cite{Sikivie:2014} that is based on the 8 T magnet currently part of the Axion Dark Matter eXperiment (ADMX)~\cite{Asztalos:2011bm}.  The estimated limit of our experiment is of the order of $10^{-14}$~GeV$^{-1}$ while the Ref.~\cite{Sikivie:2014} configuration is on the order of $10^{-16}$~GeV$^{-1}$. The main difference is that our detection system is non-superconducting and non-cryogenic, but this also means it is immune to some issues related to magnetic flux trapping, and is easier to operate, with low cost.  

\begin{figure}[t]
\centering
\includegraphics[width=3.46in]{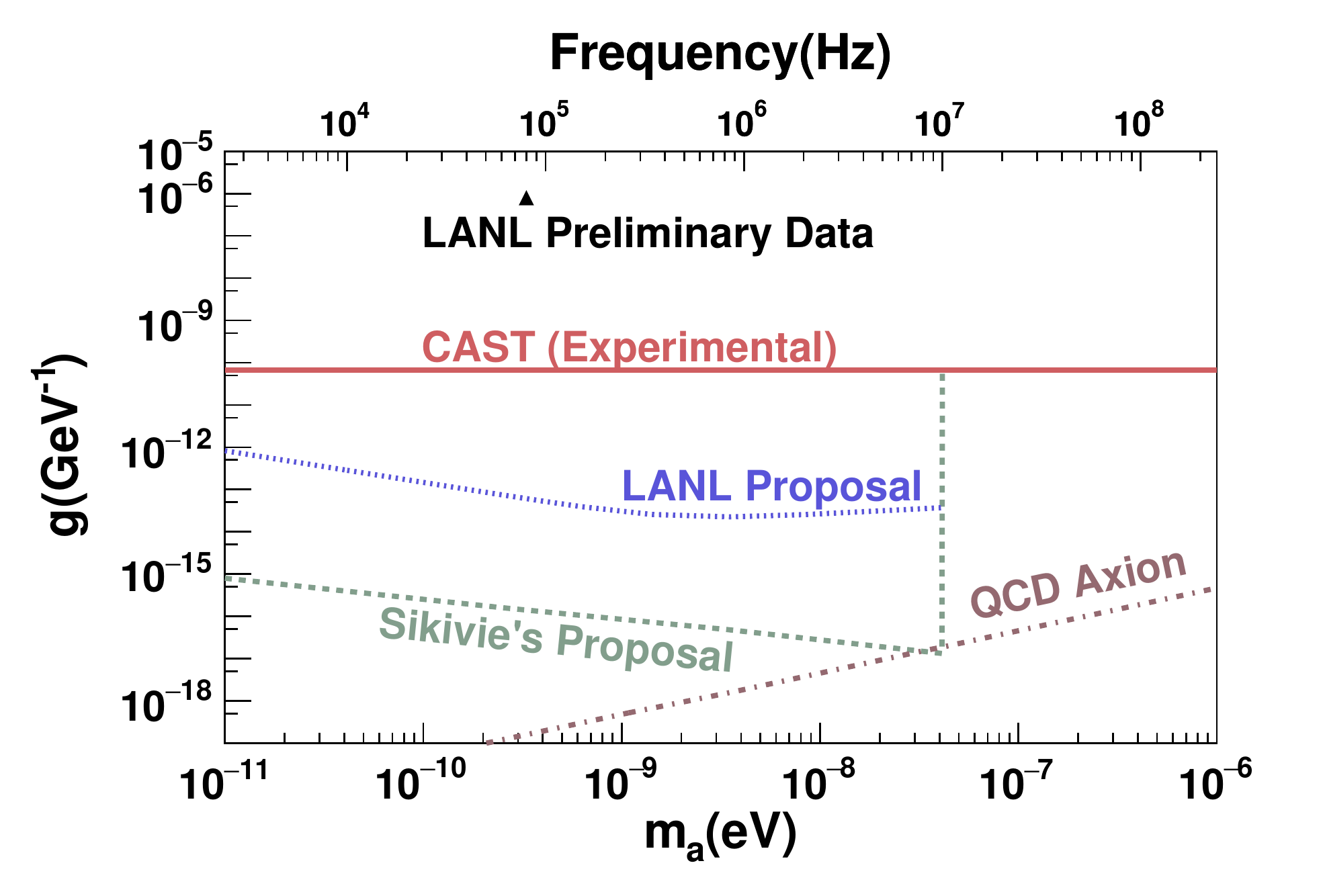}
\caption{Sensitivity to the axion-photon coupling, $g$, on the axion mass range. The region above the red line is excluded by the CAST experiment. The line labeled ``Sikivie's Proposal" is the lower sensitivity bound of one of the proposed configurations from Ref.~\cite{Sikivie:2014} using the caustic ring halo model. The purple line is the KSVZ model prediction for the QCD axion mass. Our proposed experiment is capable of searching for a  dark matter axion signal in an isothermal halo with parameters above the blue line. The triangle is the measured sensitivity of the preliminary configuration discussed in the text, which was optimized for an MRI experiment, again using the isothermal halo signal.
\label{fig:sens}} 
\end{figure}

Figure~\ref{fig:sens} shows our expected sensitivity, and that we will be able to set a new experimental limit on a significant axion mass range between 10$^{-11}$~eV and 10$^{-7}$~eV. The current best experimental bound is from the CAST experiment~\cite{CAST:NatPhys}, also shown in Fig.~\ref{fig:sens}. The current best astrophysical limits on the axion-photon coupling, from massive stars~\cite{Friedland:2012hj} and horizontal branch stars~\cite{Ayala:2014pea}, are of similar magnitude to the CAST experimental limit.  Our sensitivity estimate uses the local density, and axion signal coherence time, of the isothermal halo model \cite{Turner:1985}.  In Ref.~\cite{Sikivie:2014}, axion detection under both the isothermal halo model and the caustic ring model are considered, and the sensitivity given in Ref.~\cite{Sikivie:2014} uses the longer coherence, and higher density, of the caustic ring model \cite{Sikivie:1992,Sikivie:1996,Duffy:2008}.  Axion searches for signals predicted by the caustic ring model have increased sensitivity due to these factors, but require additional assumptions, and possible correction for the Earth's motion \cite{Duffy:2006,Hoskins:2011,Hoskins:2016,Sloan:2016}.  A thermal component to the dark matter halo, such as in the isothermal model, still occurs if high-density caustics are present near Earth, and we use this simpler assumption in our sensitivity estimate.  The KSVZ model for QCD axions \cite{Kim:1979,Shifman:1979} is included for comparison in Fig.~\ref{fig:sens}.

%MOVED FROM EARLIER - EXPAND.  See above.
%In Ref.~\cite{Sikivie:2014}, the authors described two dark matter halo models, the isothermal halo and the caustic ring halo model. A recent paper has shown that the caustic ring halo model against observations~\cite{Dumas:2015} so that in the following calculation we will use parameters of the isothermal halo model. 

Our sensitivity is mainly limited by the magnetic Johnson noise in the LC circuit. Our experiment has a lower sensitivity than the configurations proposed in Ref.~\cite{Sikivie:2014}, but this is primarily due to lower $Q$ of the LC circuit at room temperature. The upper end of the possible search range for our proposed experiment is limited by the stray capacitance in the LC circuit, as is the proposed experiment of Ref.~\cite{Sikivie:2014}: assuming the stray capacitance is 15~pF per meter and the LC circuit is in 3~m length, the maximum resonance frequency that can be reached is estimated to be $\nu<1/2\pi\sqrt{4.1~\mu\text{H}\times 45~\text{pF}}=10~$MHz, using $2\pi\nu=1/\sqrt{LC}$. 

\section{Conclusion}
We have investigated the sensitivity of a proposed experiment to search for light-mass axion dark matter using a detection system composed of an RF OPM and a LC circuit, operated at room temperature. This experiment is based on the concept initially developed by Sikivie, Sullivan, and Tanner~\cite{Sikivie:2014}, and realistic modification and optimization of an existing experiment. This experiment can explore the axion mass between $10^{-11}$~eV and $10^{-7}$~eV. The mass range is limited by the magnetic Johnson noise in the LC circuit, and the stray capacitance of the LC circuit. Our estimated sensitivity to the axion-photon coupling is up to 4 orders of magnitude better than the current best constraint~\cite{CAST:NatPhys}. Our proposed experiment can probe a significant range in the axion parameter space utilizing existing equipment such as a large-bore, 2~T magnet and an RF OPM, and represents a step forward in the search for axion dark matter that can be easily implemented in the near future.

\section{Acknowledgments}
The authors gratefully acknowledge this work was supported by the U.S. Department of Energy through the LANL Laboratory Directed Research Development Program. We would also like to thank Lisa Everett and David Tanner for helpful comments on the manuscript, and Pierre Sikivie for encouragement.

\end{document}